\documentclass[amssymb,amsmath,amsfonts,superscriptaddress,twocolumn,nopacs]{revtex4}

\usepackage{graphicx}
\usepackage{color} 
\usepackage{hyperref} 
\renewcommand{\vec}[1]{\mathbf{#1}}
\renewcommand{\d}{\mathrm{d}}

\begin{document}

\title{Tailoring the interactions between self-propelled bodies}
\author{Jean-Baptiste Caussin}
\affiliation{Laboratoire de Physique de l'\'Ecole Normale Sup\'erieure de Lyon, Universit\'e de Lyon, 46, all\'ee d'Italie, 69007 Lyon, France}
\author{Denis Bartolo}
\affiliation{Laboratoire de Physique de l'\'Ecole Normale Sup\'erieure de Lyon, Universit\'e de Lyon, 46, all\'ee d'Italie, 69007 Lyon, France}

\begin{abstract}
We classify the interactions between self-propelled particles moving at a constant speed from symmetry considerations. We establish a systematic  expansion for the two-body forces in the spirit of  a multipolar expansion. This formulation makes it possible to rationalize most of the models introduced so far within a common framework. We distinguish between three classes of physical interactions: (i)~potential forces, (ii)~inelastic collisions and (iii)~non-reciprocal interactions involving polar or nematic alignment with an induced field. This framework provides simple design rules for the modeling and the fabrication of self-propelled bodies interacting via physical interactions. A  class of possible interactions that should yield new phases of active matter is highlighted. \end{abstract}

\maketitle
\section{Introduction}
Active materials composed of motile bodies define a quickly-growing field of statistical and soft-matter physics~\cite{Vicsek2012,Marchetti2013}. Over the last decades, much attention has been devoted to the individual dynamics of  self-propelled particles (e.g.~effective diffusion and migration in an external field), and to the collective properties of large populations (e.g.~transition to collective motion, and emergence of coherent spatial patterns)~\cite{Vicsek2012,Marchetti2013,Romanczuk2012}. 

From a theoretical perspective,  the large-scale properties of active populations have been investigated for several interaction schemes at the single-particle level, see e.g.~\cite{Vicsek1995,Levine2000,Couzin2002,Peruani2008,Chate2008,Grossmann2013}. In most of the models, dynamical rules such as velocity-alignment or hard-core repulsion were included without refereeing to the microscopic physics responsible for these couplings. A priori a number of alternative phenomenological rules could be considered, yet no global framework exists to   elaborate and classify such interactions with overlooked symmetries.

From an experimental perspective, significant progress has been made over the last years, and a number of artificial active systems are now available, including self-propelled colloids~\cite{Paxton2006,Theurkauff2012,Palacci2013,Jiang2010,Bricard2013}, vibrating grains~\cite{Narayan2007,Kudrolli2008,Deseigne2010,Kumar2014}, biofilaments~\cite{Schaller2010,Sumino2012,Sanchez2012}. Now that the fabrication of motile microscopic systems is  a problem that has been partly solved, a natural next step is to consider the self-assembly of these autonomous units into new materials.  
The design of such active phases requires a deeper understanding of the interaction symmetries  between their elementary units. Surprisingly, until now, the two-body interactions  between self-propelled  colloids and filaments have been scarcely characterized.

In this paper, we  classify the symmetries of the mutual  interactions between self-propelled bodies moving at a constant speed. We first decompose  systematically  the two-body force fields in a generalized multipole expansion which solely requires that the particles live in a homogeneous space (translational invariance). We then show how little additional information about the interaction process further simplify the form of the interactions. We consider explicitly three relevant cases: (i)~ Isotropic particles interacting only via potential interactions, (ii)~Isotropic particles interacting via two-body inelastic collisions, and (iii)~Particles of arbitrary shape that interact via non-reciprocal interactions ({\em viz.} interactions that are non invariant upon Galilean transformations). We systematically exemplify our results with classical active-matter models, which we rationalize  within a unified framework. We close this paper by suggesting guidelines for the design of new active materials.

\section{Interacting self-propelled bodies: equations of motion}
By definition,  self-propelled particles convert stored internal energy to propel themselves in the absence of any external force~\cite{Marchetti2013}. We consider here the simplest framework supporting this definition:  point particles moving at a {\em constant} speed $v_0$ in a homogeneous space. More precisely, any force acting on the particle $a$ located at $\vec r_a$ and moving at a velocity $\vec v_a=v_0 \hat {\vec p}_a$  alters only its direction of motion, defined by the unit vector $\hat{\vec p}_a$.  We also assume that the particles interact via pairwise-additive couplings. The equations of motion for the particle $a$ are then the Newton's equations completed by the constant velocity constraint:
\begin{align}
\label{EM1}	\partial_t \vec r_a &= \hat{\vec p}_a ,\\
\label{EM2}	\partial_t \hat{ \vec p}_a &=  ({\mathbb I} - \hat{\vec p}_a\hat{\vec p}_a) \cdot \sum_{b\neq a}  \vec F_{b \to a}   ,
\end{align}
where $\vec F_{b \to a}$ is the force exerted by particle $b$ on particle $a$, and where we have  set $v_0=1$. The projection operator $(\mathbb{I} - \hat{\vec p}_a\hat{\vec p}_a)$ ensures that the norm of the velocity vector $\hat{\vec p}_a$ remains constant. Eq.~\eqref{EM2} implies that particles having a constant speed reorient and align along the net force that they experience. These equations, that have been extensively studied~\cite{Romanczuk2011}, can also be viewed as the asymptotic limit of a broader class of active-particle dynamics, see e.g.~\cite{Levine2000,Weber2013,Henkes2011}. Our predictions do not strictly require the propulsion speed to be a constant, but only that  some internal dissipation mechanisms causes $v_0$ to relax towards its stationary value in a time much shorter than the changes in the velocity direction. We further clarify this point and discuss the robustness of Eqs.~\eqref{EM1} and~\eqref{EM2}  in Appendix~A, by studying a specific energy-exchange model~\cite{Levine2000,D'Orsogna2006,Chuang2007}.  

\begin{figure*}
\begin{center}
\includegraphics[width=\textwidth]{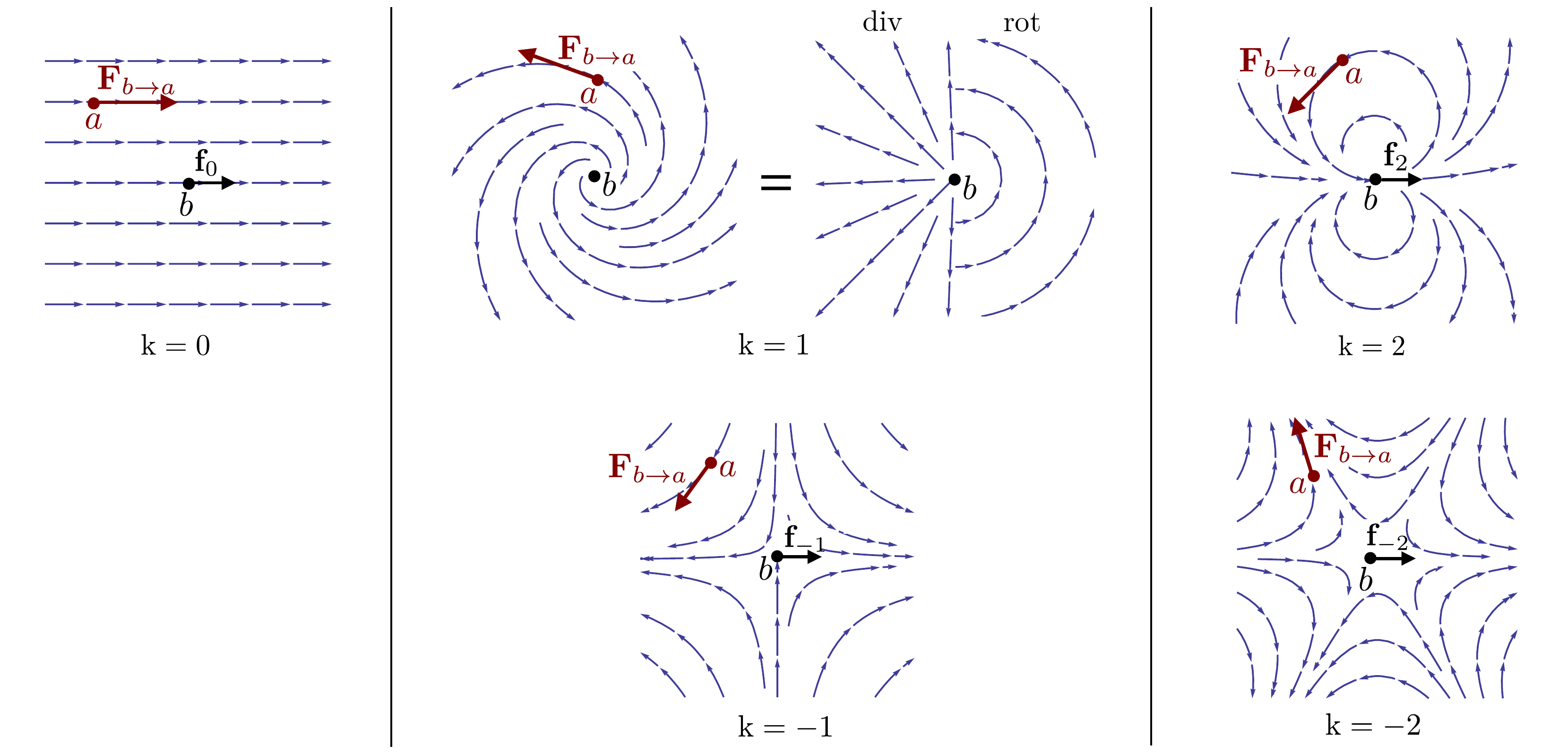}
\caption{Sketch of the first modes of the Fourier expansion Eq.~\eqref{tensor}. The force exerted by particle $b$ on particle $a$ is deduced from the field $\vec F_{b\to a}(\vec r - \vec r_b,\hat{\vec p}_a, \hat{\vec p}_b)$ at the position $\vec r = \vec r_a$. For $\rm k \neq 1$, the force field breaks the rotational symmetry, its direction is set by the vector $\vec f_{\rm k}$. We recover the symmetries of a standard multipolar expansion for $\rm k > 0$, while the $\rm k < 0$ components have negative topological charges. Note that the spiral field $\rm k=1$ is the linear superposition of a curl-free and of a divergence-less monopolar fields.}
\label{fields2D}
\end{center}
\end{figure*}

\section{Interactions in homogeneous media: Translational invariance}
We now establish a generic expression which classifies the interactions according to their angular symmetries. For  sake of clarity, we restrict the discussion to two spatial dimensions; the generalization to a 3D  system is provided in Appendix~B.
The force exerted by particle $b$ on particle $a$ is a priori a function of the two positions $\vec r_a$, $\vec r_b$ and orientations $\hat{\vec p}_a$, $\hat{\vec p}_b$. However, by definition, in a homogenous medium  $\vec F_{a\to b}(\vec r_a,\vec r_b;\hat{\vec p}_a,\hat{\vec p}_b)$ is translationally invariant, and therefore only depend on  the relative positions of the particles: $\vec r_a - \vec r_b$. We now identify 2D vectors to complex numbers, and note $\vec r_a - \vec r_b = r_{ab}  \exp(i \varphi_{ab})$, where $r_{ab}$ is the interparticle distance and $\varphi_{ab}$ the relative angular position.
Without any additional assumption, we Fourier transform $\vec F_{b \to a}$ with respect to $\varphi_{ab}$ and obtain:
 \begin{align}
 \label{Fourier}
 	\vec F_{b \to a} = & \sum_{\rm k} f_{\rm k}(r_{ab}, \hat{\vec p}_a, \hat{\vec p}_b) \,  \mathrm e^{i \left[ \mathrm k \varphi_{ab} + \psi_{\rm k}(r_{ab}, \hat{\vec p}_a, \hat{\vec p}_b) \right]}.
 \end{align}
Transforming the real and imaginary parts into polynomial series in $\cos \varphi_{ab}$ and $\sin \varphi_{ab}$, and after some elementary algebra, Eq.~\eqref{Fourier} is recast into a more intuitive   expansion akin to a multipolar series:
\begin{align}
\label{tensor}
	\vec F_{b \to a} &= \vec f_0 + f_1^{\rm div} \hat{\vec r}_{ab} + f_1^{\rm rot} \boldsymbol \epsilon \cdot \hat{\vec r}_{ab}\\ & +  f_{-1} \left(2 \frac{\vec f_{-1}  \vec f_{-1}}{f_{-1}^2} - {\mathbb I}\right) \cdot \hat{\vec r}_{ab} + \vec f_2  \cdot (2 \hat{\vec r}_{ab} \hat{\vec r}_{ab} - {\mathbb I}) + \ldots\nonumber
\end{align}
 $f_1^{\rm div} = f_1 \cos \psi_1$ and $f_1^{\rm rot} = - f_1 \sin \psi_1$ are scalars, $\boldsymbol \epsilon$ is the completely antisymmetric Levy-Civita symbol, and 
all the other coefficients are vectors: $\vec f_0 = f_0(\cos \psi_0, \sin \psi_0)$, $\vec f_{-1} = f_{-1} (\cos (\psi_{-1}/2),\sin (\psi_{-1}/2))$ and $\vec f_2 = f_2(\cos \psi_2, - \sin \psi_2)$.
 We recall that all the $\vec f_k$  depend in principles on the relative position and on the orientations of particles $a$ and $b$. The terms in Eq.~\eqref{tensor} are classified according to their angular periodicity, the index $\rm k$ corresponds to the topological charge of the force field induced by the particle $b$: the larger $\rm k$, the faster the angular variations of the field.
Eq.~\eqref{tensor} is already a pivotal result of this paper as it describes all the possible symmetries of the field in which a self-propelled particle reorients.  In order to gain more physical insight into this formal expansion, the first modes are plotted in Fig.~\ref{fields2D}, and discussed below.
\begin{itemize}
\item[\textbullet] The $\rm k = 0$ component is a field having a constant orientation. The particle $a$ reorients along $\vec F_{a \to b}$, which is aligned with the vector $\vec f_0(r_{ab},\hat{\vec p}_a,\hat{\vec p}_b)$ regardless of the relative positions of the particles.  
\item[\textbullet] The $\rm k = 1$ component in Eq.~\eqref{tensor} generically corresponds to a spiral force field. To better understand its contribution, it is convenient to distinguish between its divergence and rotational components as done in Eq.~\ref{tensor} and Fig. 1. The first term, $f_1^{\rm div} \hat{\vec r}_{ab}$, has the symmetry of a monopolar field. This term therefore gives rise to repulsive (resp. attractive) couplings.  For example, particle $a$ reorients along $\hat{\vec r}_{ab}$ if $f_1^{\rm div}(r_{ab},\hat{\vec p}_a,\hat{\vec p}_b)  > 0$. 
The second term, the rotational part $f_1^{\rm rot} \boldsymbol \epsilon \cdot \hat{\vec r}_{ab}$, breaks the bottom-top symmetry in 2D, which implies that $f_1^{\rm rot}$ is a pseudoscalar quantity. Consequently, $f_1^{\rm rot}$ is non-zero only for particles having some chiral features. In such an interaction field, a chiral particle $b$ could be forced to circle around its neighbor.  
\item[\textbullet] The $\rm k = -1$ component has the form of an hyperbolic elongation field with a negative topological charge.  The $\rm k = 2$ contribution has the dipolar symmetry, the direction of the dipole being set by the vector $\vec f_2(r_{ab},\hat{\vec p}_a,\hat{\vec p}_b)$. Similarly, the higher-order terms of the Fourier expansion correspond to force fields having positive or negative topological charges. Note that the contributions from the $\rm k<0$ components do not correspond to  conventional multipoles associated with a Laplacian field.

\end{itemize}
We have classified all the possible symmetries of the fields that cause the particle reorientation. In order to fully prescribe the orientational dynamics, we now have to specify how the $\vec f_{\rm k}$s relate to the particle orientations and relative distance. In order to do so, we  focus on three types of interactions which encompass most of the numerical and experimental systems.

\section{Potential interactions between isotropic particles} We first consider the simplest  possible setup:  isotropic particles interacting via potential interactions. $\vec F_{a \to b} = -\nabla_a U(r_{ab})$ derives from a potential $U(r_{ab})$ which only depends on the interparticle distance. Consequently, $\vec F_{a \to b}$ readily reduces to the sole curl-free part of the  mode $\rm k = 1$ in Eq.~\eqref{tensor}. The force field has the symmetry of a monopole, with a strength $f_1^{\rm div}(r_{ab}) = -\partial_{r_{ab}} U(r_{ab})$, it  results in attractive or repulsive couplings. Such potential interactions have been studied in a number of numerical models, see e.g.~\cite{Levine2000,D'Orsogna2006,Chuang2007,Hanke2013,Ferrante2013}. For instance, d'Orsogna {\em et al.} demonstrated that  forces deriving from a Morse potential can lead to the emergence of a number of patterns all having an a rotational symmetry, such as vortices, rings and circular clumps~\cite{D'Orsogna2006}.
We stress that potential forces cannot explicitly couple the  velocities of isotropic particles.   However we will show in Sect.~\ref{non-reciprocal}, and in Appendix~C, that potential forces can yield net alignment interactions between slender bodies. 

\section{Inelastic collisions between isotropic particles} We now turn to a more general situation and assume that the particles undergo physical two-body collisions, where  $\vec F_{a\to b}$ could also be associated to a dissipative process. However, we still restrain here  to the situation where $\vec F_{a\to b}$ is  invariant upon Galilean transformations. In other words, $\vec F_{a\to b}$ is assumed to only depend on the relative position, $\vec r_{ab}$, and on the relative orientation/velocity, $\hat{\vec p}_a - \hat{\vec p}_b$. The coefficients of the Fourier series hence depend only on $\hat{\vec p}_a - \hat{\vec p}_b$, and the vector coefficients $\vec f_{\rm k \neq 1}$ in Eq.~\eqref{tensor} are all oriented along $\hat{\vec p}_a - \hat{\vec p}_b$:  
\begin{equation}
\vec f_{\rm k} = f_{\rm k}(r_{ab},|\hat{\vec p}_a - \hat{\vec p}_b|) \, (\hat{\vec p}_a - \hat{\vec p}_b).
\end{equation}
It follows from Eq.~\eqref{tensor}   that $\vec F_{b \to a} = - \vec F_{a \to b}$.  Even though the self-propulsion mechanism does not conserve momentum, see Eq.~\eqref{EM1}, the  invariance of the forces upon Galilean transformations implies that they obey the Newton's third law. This situation typically corresponds to the model for vibrated polar disks introduced in~\cite{Weber2013}. Weber {\em et al.} studied numerically a system of polar grains set in motion by the vibration of a substrate and interacting via short-range inelastic collisions~\cite{Narayan2007,Kudrolli2008,Deseigne2010,Grossman2008}. They model the interactions by the spring-and-dash-pot model and assume that $\vec F_{a \to b} = -\lambda [ (\hat{\vec p}_a - \hat{\vec p}_b)\cdot \hat{\vec r}_{ab} ]\hat{\vec r}_{ab}+ \kappa (d - r_{ab}) \hat{\vec r}_{ab}$. The first term is associated with viscous friction, and the last term corresponds to elastic repulsion for $r_{ab} < d$. They observed numerically that such inelastic collisions can lead to velocity alignment interactions, thereby giving rise to a macroscopic polar order. In the classification given by Eq.~\eqref{tensor}, the force corresponds to a combination of the three components $\rm k = 0$, 1 and 2, where $\vec f_0 = \vec f_2 = - (\lambda/2) (\hat{\vec p}_a - \hat{\vec p}_b)$ and $f_1^{\rm div} = \kappa (d - r_{ab})$. Beyond this specify model which  beautifully accounts for experimental results, the present framework makes it possible to provide a clear design rule for collision-induced velocity alignment at the two-particle level.  The lowest-order mode of $\vec F_{a\to b}$, $\vec f_0 = f_0(r_{ab},|\hat{\vec p}_a - \hat{\vec p}_b|) \, (\hat{\vec p}_a - \hat{\vec p}_b)$, is the only one that promotes a net polar alignment  regardless of the two particles conformation provided that $f_0 < 0$ ($f_0 > 0$ would result in anti-alignement interactions). Particles interacting via $\vec f_0$  evolve according to
\begin{equation}
\partial_t \hat{\vec p}_a = - f_0(r_{ab},|\hat{\vec p}_b - \hat{\vec p}_a|) \, ({\mathbb I}- \hat{\vec p}_a\hat{\vec p}_a) \cdot \hat{\vec p}_b.
\end{equation}
 Consequently, any other  contribution to the force expansion, Eq.~\eqref{tensor}, would compete with this velocity-alignment rule and, in principle, could yield macroscopic states with more complex symmetries that a mere polar phase. 
\section{Nonreciprocal interactions.}
\label{non-reciprocal}

\begin{figure}
\begin{center}
\includegraphics[width=\columnwidth]{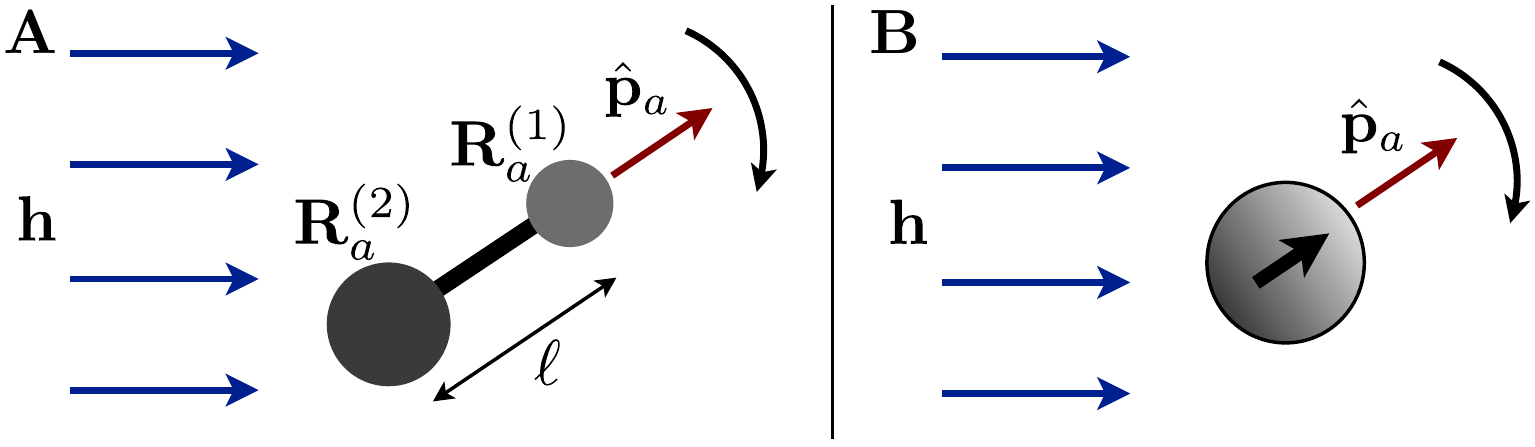}
\caption{Polar alignment in an external field $\vec h$. A-- An asymmetric dumbbell reorients in the field. B-- An isotropic particle carrying a polar internal structure can also align its velocity with the field.}
\label{fields2D}
\end{center}
\end{figure}
We now consider forces that are not necessarily invariant upon Galilean transformations, and therefore relax the reciprocal condition. $\vec F_{a\to b}$  can now be different from $-\vec F_{b\to a}$. 
Prominent instances are: (i) hydrodynamic interactions between simmers~\cite{Saintillan2013,Brotto2013}: when swimming in a liquid, particle $b$ creates a flow field that causes particle $b$ to reorient in this flow, thereby  yielding effective long-range interactions. (ii) Polar phoretic particles respond to the local variations of a scalar quantity (chemical potential~\cite{Paxton2006,Theurkauff2012,Aranson2013}, temperature \cite{Jiang2010}). They are also prone to align in the phoretic field induced by their neighbors~\cite{Bickel2014,Saha2013}. 
More generally, we here describe particles that reorient in a field originated from its neighbors. Note that this broader class encompasses the  potential interactions that we first considered. The field in which the particle aligns would precisely correspond to the gradient of the potential.

In all that follows, we denote by $\vec h(\vec r - \vec r_b, \hat{\vec p}_b)$ the field created by particle $b$ at position $\vec r$. We assume that $\vec h$ depends on time only via the particles' conformation (this hypothesis corresponds to a zero-Peclet number approximation in the context of phoretic particles) and that it does not dependent on $\vec p_a$. This approximation is well-suited in a far-field description, since at long distance the orientation of particle $a$ does not modify the field induced by particle $b$.

We now have to specify how the particle reorients in the external field $\vec h$ to fully define the expression of $\vec F_{a\to b}$. In view of gaining  more physical intuition, we first introduce a  prototypical  dumbbell model that can be simply solved.

\subsection{Self-propelled dumbbells} 
We consider dumbbells made of two rigidly connected disks separated by a distance $\ell$, Fig. 2A. Looking at the motion of the dumbbell $a$, we denote by $\vec R_a^{(1)}$ and $\vec R_a^{(2)}$ the positions of the two disks. We assume that the particle propels along its main axis: $\hat{\vec p}_a = (\vec R_a^{(1)} - \vec R_a^{(2)})/\ell$. The dynamics of the dumbbell is then modeled as follows. Both disks respond linearly to the field $\vec h(\vec R - \vec r_b,\hat{\vec p}_b$) exerted by particle $b$, where $\vec R = \vec R_a^{(j)}$, $j = 1,2$. As thoroughly demonstrated in Appendix~C, the equation of motion  of the dumbbell orientation then reduces to:
\begin{align}
\label{dumbbelltext}	&\partial_t \hat{\vec p}_a = (\mathbb{I} - \hat{\vec p}_a \hat{\vec p}_a) \cdot \left[ \alpha \vec h(\vec r_a - \vec r_b,\hat{\vec p}_b) + \beta (\hat{\vec p}_a \cdot \nabla) \vec h  \right],
\end{align}
where $\alpha$ and $\beta$ are two constant scalar quantities.
This equation correspond to Eq.~\eqref{EM2} with $\vec F_{a \to b} $ given by:
\begin{equation}
	\vec F_{a \to b} = \alpha \vec h(\vec r_a - \vec r_b,\hat{\vec p}_b) + \beta (\hat{\vec p}_a \cdot \nabla) \vec h + \mathcal O(\ell^2 \nabla^2) .
\end{equation}
We also show in Appendix~C that $\alpha$ can either be a positive or a negative quantity depending of the relative mobility coefficients of the two disks.  Hence two different behaviors are obtained. If the two disks are {\em not} identical (polar dumbbell), $\alpha \neq 0$ and the force is $\vec F_{a \to b} = \alpha \vec h(\vec r_a - \vec r_b,\hat{\vec p}_b) + \mathcal O(\ell \nabla)$. The dumbbell aligns with, or opposite to $\vec h(\vec r_a - \vec r_b,\hat{\vec p}_b)$. Conversely, if the two disks are identical, then $\alpha = 0$ and the first term in Eq.~\eqref{dumbbelltext} vanishes. The force then reduces to $\vec F_{a \to b} =  \beta (\hat{\vec p}_a \cdot \nabla) \vec h$, and the dumbbell aligns nematically in a direction set by the field gradient. This minimalist setup already shows that multiple field alignment rules can exist. We now go beyond this specific picture and discuss more generally the polar and nematic cases. 
\subsection{Polar alignment in an induced field} 
We first assume polar alignment in the direction of the field, which, regardless of the shape of the particle, translates into:
\begin{equation}
\label{polar}
	\vec F_{a \to b} = \alpha \, \vec h(\vec r_a - \vec r_b, \hat{\vec p}_b) ,
\end{equation}
where $\alpha$ is here a phenomenological  coefficient. The force is independent of $\hat{\vec p}_a$. As a consequence, the coefficients of the  expansion in Eq.~\eqref{tensor}  only depend on $r_{ab}$ and $\hat{\vec p}_b$, and the scalar coefficients $f_1^{\rm div}$ and $f_1^{\rm rot}$ solely depend on the interparticle distance. In addition, the vector coefficients $\vec f_{\rm k \neq 1}$ are necessarily oriented along $\hat{\vec p}_b$: 
\begin{equation}
\vec f_{\rm k} = f_{\rm k}(r_{ab}) \hat{\vec p}_b.
\end{equation}
 We  illustrate this result with three concrete examples.
(i)~This situation has been considered in the context of interacting polar swimmers in confinement~\cite{Brotto2013,Lefauve2014}, and of the motion of biofilament in plant cytoskeleton~\cite{Woodhouse2013}. For confined swimmers, Eq.~\eqref{tensor} reduces to the sole dipolar  term ($\rm k=2$) which reflects the potential flow induced by any type of self-propelled object in a rigidly confined liquid film. In a different context,  Kumar {\em et al.} have demonstrated experimentally and numerically that  pointy rods lying on a shaken bed of isotropic grains experience self-propulsion and velocity alignment interactions~\cite{Kumar2014}. The  rods interact effectively as they align locally with  the polar displacement field induced by the motion of their neighbors on top of the shaken bead layer.
(ii)~This type of coupling is also responsible for the emergence of flocking patterns and of spontaneously flowing phases in the ensembles of colloidal rollers introduced in~\cite{Bricard2013}. The  velocity of a roller aligns in the direction of the flow field induced by  the surrounding motile colloids.  As show in~\cite{Bricard2013} (supplementary materials), within a far-field approximation, the roller-roller interaction combines the first terms of Eq.~\eqref{tensor}, $\rm k = 0$, 1 and 2. We emphasize that the colloidal rollers have a perfectly isotropic shape, however their velocity very quickly relaxes in the direction of their dipolar electric-charge distribution. This polar internal degree of freedom is responsible for the polar-alignement rule in the external field, see Fig.~2B.  (iii)~Finally this type of coupling also encompasses the interactions used in  agent-based models for collective motion, provided that they involve pairwise-additive interactions. In all these sequels of the seminal Vicsek model~\cite{Vicsek1995}, see e.g.~\cite{Peruani2008,Farrell2012,Bertin2009}, the interactions exactly correspond to the first Fourier mode, $\vec f_0 = f_0(r_{ab}) \hat{\vec p}_b$.
On top of this velocity-alignment rule, short-range repulsion and long-range attraction have also been considered to reproduce the morphology of cohesive flocks akin to animal populations~\cite{Couzin2002,Chate2008,Gregoire2004}. They are associated with the $\rm k = 1$ component of the multipole expansion. 

\subsection{Nematic alignment in an induced field}   
In line with the previous discussion, and with Eq.~\eqref{dumbbelltext}, some motile particle, say $a$, can also align nematically with the  field $\vec h(\vec r_a-\vec r_b,\hat {\vec p}_b)$ induced by particle $b$. This condition translates into the following generic expression for the induced force:
\begin{equation}
\label{nematic}
	\vec F_{a \to b} = \beta \, \vec N(\vec r_a-\vec r_b, \hat{\vec p}_b) \cdot \hat{\vec p}_a ,
\end{equation}
were $\vec N$ is a tensor  that can be constructed from the field $\vec h$ and the $\nabla$-operator, depending on the type of induced field and particle shape we are talking about. For instance, for the symmetric dumbbells $\vec N$ takes the simple form: $\vec N=\nabla \vec h$, see Eq.~\eqref{dumbbelltext}. More generally, when the nematic-alignment rule arises from the slenderness of the particle, the expression of the tensor $\vec N$ is given by the so-called Jeffery's equation first introduced in the context of fluid mechanics~\cite{Jeffery1922,Saintillan2008}.  Eq.~\eqref{nematic} implies that the force $\vec F_{b \to a}$ linearly depends on $\hat{\vec p}_a$. The amplitudes of the Fourier coefficients, $f_{\rm k}$ in Eq.~\eqref{Fourier}, therefore include a factor $\hat{\vec p}_a \cdot \hat{\vec r}_{ab}$ or $\hat{\vec p}_a \cdot \hat{\vec p}_b$, and the vectors $\vec f_{\rm k \neq 1}$ in Eq.~\eqref{tensor} are oriented along $\hat{\vec p}_b$. The k-coefficient of the classification in Eq.~\eqref{tensor} takes the generic form:
\begin{equation}
\vec{f}_k=\left[f_{k}^{(pp)}(r_{ab})\hat{\vec p}_a \cdot \hat{\vec p}_b+f_{k}^{(rp)}(r_{ab})\hat{\vec p}_a \cdot \hat{\vec r}_{ab}\right]\hat{\vec p}_b .
\end{equation} 

An important example concerns the zeroth-order term: both the $f_0^{pp}$ and the  $f_0^{rp}$ contributions promote nematic alignment between the two interacting particles $a$ and $b$ for this first lowest-order term. We are now aware of any experimental realization, where self-propelled particles would experience nematic interactions mediated by an induced field. For instance in the seminal pusher/puller model of Saintillan et al. Active swimmer align nematically in the far-field flow induced by their motion. However these hydrodynamic interactions only correspond to the $\rm k=+3$ mode in Eq.~\eqref{tensor}, thereby yielding complex spatiotemporal fluctuations in swimmer suspensions which cannot support nematic order~\cite{Saintillan2008}.

\subsection{Higher-order symmetries} 
For particles having more complex shapes, the reorientation in the field in principle involves higher symmetries than the polar or nematic modes. For instance, the force can include higher-order terms such as $\vec F_{a \to b} = \gamma \, \vec M(\vec r_a-\vec r_b, \hat{\vec p}_b)\! : \! \hat{\vec p}_a \hat{\vec p}_a$, where $\vec M$ is a third-rank tensor build from the field $\vec h$ and its derivatives. Following the same procedure, classifying the interactions associated with these higher-order couplings is straightforward.


\section{Discussion}

Building only on symmetry arguments, we have introduced a formal classification of all the possible interactions between self-propelled particles. This classification does not only rationalize all the previously introduced models within a unique formal framework, it also brings two unanticipated perspectives on the physics of interacting motile bodies.

Firstly, from a technical perspective, the classification defined by Eq.~\eqref{EM2} makes it possible to quickly identify the salient features of the two-body dynamics, even if the interactions take a complex form. For instance, it offers a simple mean to check wether the interactions promote velocity alignment (polar or nematic).   In  practical terms,  Eqs.~\eqref{EM2} and \eqref{tensor} imply that that $\vec F_{b\to a}$ supports a net alignment of the velocities only if  its zero mode  does not vanish. Equivalently, a simple criteria for two-body alignment is that the angular integral of the two-body force does not vanish. From Eq.~\eqref{Fourier}, it necessarily takes the simple form:
\begin{equation}
\int\! \vec F_{b\to a} \, {\rm d}\varphi_{ab}=\epsilon\hat{\vec p}_b+\epsilon'\hat{\vec p}_a
\label{angularintegral}
\end{equation}
where the $\hat{\vec p}_a$ term does not contribute to the orientational dynamics, see Eq.~\eqref{EM2}. Any nonzero value for $\epsilon$ yields alignment. To distinguish between interactions that promote polar or nematic ordering, one should also look at the dependence on $\hat{\vec p}_a$ of $\epsilon$. If $\epsilon$ is unchanged as the particle $a$ changes its orientation ($\hat{\vec p}_a\to-\hat{\vec p}_a$), then $\vec F_{a\to b}\sim \hat{\vec p}_b$ and  leads to polar alignment in Eq~\eqref{EM2}. In contrast, if the sign of $\epsilon$ changes as  particle $a$ changes its orientation ($\hat{\vec p}_a\to-\hat{\vec p}_a$), then $\vec F_{a\to b}$ leads to nematic alignment in Eq~\eqref{EM2}.

Secondly, going back to the initial motivation of this work, the phase of an ensemble of interacting particles is set by the symmetries of the interactions at the microscopic level. Although this statement is obvious for equilibrium systems, it has been surprisingly overlooked when considering active matter. Until now, all the theoretical models and the (quantitatively characterized) experimental realizations of active matter involving physical interactions have been restricted to the $\rm k=0,1,2,3$ modes of the classification~\eqref{tensor}. Eq.~\eqref{tensor} demonstrate that there exist a number of possible interactions rules between self-propelled particles that have not been considered at all, even though they should yield novel macroscopic phases of active matter. A special attention should be devoted to the $\rm k=-1$ and $\rm k=-2$ modes of Eq.~\eqref{tensor}. All the kinetic theories of active matter confirmed that interactions with a low angular symmetry strongly contribute to the large-scale hydrodynamics of these systems. These  types of interactions are  therefore expected to significantly contribute to the phase behavior of unanticipated active materials.

\begin{acknowledgments}
This work was partly funded by ANR grant MITRA and by the Institut Universitaire de France.
\end{acknowledgments}

\appendix
\section{Impact of small fluctuations of the particle speed}
\label{appendix_SPP}

In this section, we further extend the range of validity of the equations of motion \eqref{EM1}--\eqref{EM2}. Let us consider the more general case where small fluctuations of the particle speed are allowed. Following e.g.~\cite{D'Orsogna2006,Levine2000}, we model self-propulsion by a non-linear friction force and assume the following equations of motion:
\begin{align}
\label{EM_friction1}	\partial_t \vec r_a &= \vec v_a ,\\
\label{EM_friction2}	\partial_t \vec v_a &= \frac{1}{\tau} \left( 1 - | \vec v_a | \right)\vec v_a + \sum_{b \neq a} \vec F_{b \to a} + \boldsymbol \xi_a (t) .
\end{align}
 Eq.~\eqref{EM_friction2} is a momentum conservation equation. The first term on the r.h.s.~models self-propulsion. It describes the conversion of internal energy into translational motion, and accounts for dissipative friction forces. As a result, the particle speed relaxes to $|\vec v| = 1$. In addition, $\sum \vec F_{b\to a}$ is the total external force acting on particle $a$. We also include here a possible noise term $\boldsymbol \xi_a(t)$ with zero mean and variance $\langle \boldsymbol \xi_a(t) \boldsymbol \xi_a(t') \rangle = 2 D \delta(t-t') {\mathbb I}$ . For the sake of clarity we discard active noise, which would stem from fluctuations in the propulsion mechanism and could be easily included as well, see e.g.~\cite{Romanczuk2011}.

We focus on small fluctuations of the speed, i.e.~we assume that $\tau D = \mathcal O(\epsilon)$ and $\tau F_{b\to a} = \mathcal O(\epsilon)$, where $\epsilon \ll 1$. It is worth noting that Eq.~\eqref{EM_friction2} is the only possible expression for the propulsion force at first order in $\epsilon$.  For  sake of simplicity we restrain here to particles moving in a 2D space, as the  generalization to 3D is straightforward.

In order to obtain stochastic equations for the particle position and orientation, we have to integrate out the speed fluctuations. This is easily done by  using polar coordinates. We introduce $\vec v_a =(1+u_a) \hat{\vec p}_a$, where $u_a = \mathcal O(\epsilon)$, and we describe the particle orientation by the polar angle $\theta_a$: $\hat{\vec p}_a = (\cos \theta_a, \sin \theta_a)$. Similarly, we write the force as: $\sum_b \vec F_{b\to a} \equiv F_a(t) \left( \cos \phi_a(t), \sin \phi_a(t) \right)$. Eqs.~\eqref{EM_friction1}--\eqref{EM_friction2} are recast into:
\begin{align}
\label{eq_ra}	\partial_t \vec r_a &= (1+u_a) \hat{\vec p}_a ,\\
\label{eq_u}	\tau \partial_t u_a &= -u_a + \tau D + \tau F_a \cos(\theta - \phi_a) +  \xi_a^u (t) ,\\
\label{eq_theta}	\tau \partial_t \theta_a &= - \tau F_a \sin(\theta - \phi_a) + \tau  \xi_a^\theta (t) + \mathcal O(\epsilon^2) ,
\end{align}
where $\xi_a^u$ and $\xi_a^\theta$ are independent white noises with zero mean and variance $\langle \xi_a^u(t) \xi_a^u(t') \rangle = \langle \xi_a^\theta(t) \xi_a^\theta(t') \rangle = 2 D \delta(t-t')$. The term $\tau D$ in Eq.~\eqref{eq_u} is a spurious drift term that classically results from  the Stratonovich discretization scheme used here to define the noise terms, see e.g.~\cite{Romanczuk2012}.

Eqs.~\eqref{eq_u}--\eqref{eq_theta} involve two well separated time scales. In Eq.~\eqref{eq_u}, the particle speed relaxes to $|\vec v| = 1$ in a time $\sim \tau$. The force and noise terms only give rise to subdominant corrections. Conversely, in Eq.~\eqref{eq_theta} the particle orientation evolves on a much longer time scale $\sim \tau/\epsilon$. Indeed, self-propulsion corresponds to a spontaneous breaking of the rotational symmetry, the orientation is not constrained to relax toward a position set by any potential. Only the external forces and noise dictate the orientational dynamics.

We now want to describe the particle dynamics on time scales much longer than $\tau$. In other words, 
we want to average the dynamics over the fast variations of the particle speed. Firstly, we integrate Eq.~\eqref{eq_u} and combine this result with Eq.~\eqref{eq_ra}, which yields
\begin{align}
	&\partial_t \vec r_a = (1+\tau D) \hat{\vec p}_a + \tau \xi_t(t) \, \hat{\vec p}_a  \nonumber\\&\hspace{0.3cm}+ \sum_b\! \int \! G(t-t') \, \tau \, [\vec F_{b\to a}(t') \cdot \hat{\vec p}_a(t')] \hat{\vec p}_a(t)\, \d t' ,
\end{align}
and
\begin{equation}
	\partial_t \theta_a = -F_a \sin(\theta - \phi_a) + \xi_a^\theta (t) ,
\end{equation}
where $\xi_t(t)$ is a colored translational noise, with zero mean and correlations defined by $\langle \xi_t(t) \xi_t(t') \rangle = (D/\tau) \exp{\left[ -| t-t' |/\tau \right]}$. The kernel $G$ is given by $G(t) = \tau^{-1} \exp{[-(t)/\tau}] \, \Theta(t)$, where $\Theta$ is the Heaviside step function. When considering only time variations at scales much larger than $\tau$, the time correlations of the translational noise vanish, $\langle \xi_t(t) \xi_t(t') \rangle \sim 2D \delta(t-t')$ and we recover a simple Markovian kernel $G(t-t') \sim \delta(t-t')$.
Coming back to vector notations for the particle orientation, the two coarse-grained equations of motion are:
\begin{align}
\label{eq_pos}	\partial_t \vec r_a &= \hat{\vec p}_a + \tau \sum_b \vec F_{b \to a} \cdot \hat{\vec p}_a \hat{\vec p}_a+ \tau \xi_t(t) \, \hat{\vec p}_a ,\\
	\partial_t \hat{ \vec p}_a &= ({\mathbb I} - \hat{\vec p}_a\hat{\vec p}_a) \cdot \left[ \sum_b \vec F_{b \to a}(t) + \boldsymbol \xi_a (t) \right] .
\end{align}
We  recover Eq.~\eqref{EM2} for the orientation dynamics. The velocity of the particle, Eq.~\eqref{eq_pos}, is the sum of the self-propulsion $\hat{\vec p}_a$ and additional advection terms due to the interactions and the noise. However, these contributions correspond to small corrections as $\tau F_{b\to a}$ and $\tau D$ were assumed to be of order $\epsilon$. The equations of motion that we assumed in the main text therefore correspond to the limit $\tau \to 0$ of this more general model. In other words,  Eqs.~\eqref{EM1} and~\eqref{EM2} are valid when the relaxation of the speed towards $|\vec v| = 1$ is much faster than the modification of the particle's direction due to interactions and noise.
\section{Generalization to 3D}
\label{appendix_3D}

The 3D generalization of Eq.~\eqref{Fourier} is a spherical harmonics expansion, and contains a much larger number of terms. However, most of them are discarded by the symmetries of the interactions. For potential interactions between isotropic particles, the force is radial. For Galilean-invariant interactions, the force field depends on a unique vector, $\hat{\vec u} = (\hat{\vec p}_b - \hat{\vec p}_a)/|\hat{\vec p}_b - \hat{\vec p}_a|$. Its expression is therefore invariant by rotation around the $\hat{\vec u}$-axis. For particles aligning in a field, the force field also depends on a unique vector, $\hat{\vec u} = \hat{\vec p}_b$, if the alignment rule is polar. The case of particles aligning nematically is readily inferred from the polar case by adding a factor $\hat{\vec p}_a \cdot \hat{\vec p}_b$ or $\hat{\vec p}_a \cdot \hat{\vec r}_{ab}$ in the expression of the force.

These symmetry requirements greatly simplify the spherical harmonics expansion. In an orthonormal basis $(\hat{\vec x}, \hat{\vec y}, \hat{\vec u})$, we find:
\begin{align}
	\vec F_{b \to a} = & f_0 \hat{\vec u}
+
	\sum_{\rm k \geq 1} \left[ f_{\rm k} \left|
\begin{array}{ll}
	Y_{\ell,x}(\hat{\vec r}_{ab})\\
	Y_{\ell,y}(\hat{\vec r}_{ab})\\
	Y_{\ell,0}(\hat{\vec r}_{ab})
\end{array}
\right.
+
f_{\rm -k} \left|
\begin{array}{ll}
	-Y_{\ell,x}(\hat{\vec r}_{ab})\\
	-Y_{\ell,y}(\hat{\vec r}_{ab})\\
	Y_{\ell,0}(\hat{\vec r}_{ab})
\end{array}
\right.
 \right.\nonumber \\
&\left.
\hspace{2cm}+ \Omega_{\rm k} \left|
\begin{array}{ll}
	-Y_{\ell,y}(\hat{\vec r}_{ab})\\
	Y_{\ell,x}(\hat{\vec r}_{ab})\\
	0
\end{array}
\right.
 \right] ,
\end{align}
where $Y_{\ell,m}$ are the spherical harmonics, $Y_{\ell,x} = (Y_{1,-1}-Y_{1,1})/2$ and $Y_{\ell,y} = -(Y_{1,-1}+Y_{1,1})/(2i)$. In vector notations, the above equation takes a much more compact form:
\begin{align}
\label{expansion3D}
	\vec F_{b \to a} &= f_0 \hat{\vec u} + f_1 \hat{\vec r}_{ab} + f_{-1} (2 \hat{\vec u} \hat{\vec u} - {\mathbb I})\cdot \hat{\vec r}_{ab} + \Omega_1 \hat{\vec u} \times \hat{\vec r}_{ab} \nonumber\\&\hspace{0.5cm} + f_2 \hat{\vec u}  \cdot (3 \hat{\vec r}_{ij} \hat{\vec r}_{ij} - {\mathbb I}) + ...
\end{align}
The components $f_{\rm k > 0}$ correspond to a standard multipolar series, they have the symmetries of monopolar, dipolar, quadrupolar fields etc. The coefficients $f_{\rm -1}$ and $\Omega_{\rm k}$ are associated with elongational and rotational fields, respectively.
\section{Equations of motion for a dumbbell-shaped particle}
\label{appendix_dumbbell}
 We consider particles composed of two rigidly connected disks separated by a fixed distance $\ell$. Looking at the motion of particle $a$, we denote by $\vec R_a^{(1)}$ and $\vec R_a^{(2)}$ the positions of the two disks. We assume here that the dumbbell propels along its principal axis: $\hat{\vec p}_a = (\vec R_a^{(1)} - \vec R_a^{(2)})/\ell$. The dynamics of the dumbbell is modeled as follows. We assume that the disk $j$ propels itself in the direction $\hat{\vec p}_a$ and experiences the force field $\vec h(\vec R_a^{(j)} - \vec r_b,\hat{\vec p}_b$) exerted by particle $b$ ($j = 1$ or 2). The corresponding equations of motion are:
\begin{align}
	&\partial_{tt} \vec R_a^{(1)} = \frac{1}{\tau_1} \left( \hat{\vec p}_a - \partial_t \vec R_a^{(1)} \right) + \vec h(\vec R_a^{(1)} - \vec r_b,\hat{\vec p}_b) + \vec F_a^{(2)\to(1)} ,\\
	&\partial_{tt} \vec R_a^{(2)} = \frac{1}{\tau_2} \left( \hat{\vec p}_a - \partial_t \vec R_a^{(2)} \right) + \vec h(\vec R_a^{(2)} - \vec r_b,\hat{\vec p}_b) + \vec F_a^{(1)\to(2)},
\end{align}
where, the first term on the r.h.s accounts for the competition between the propulsion and the drag force experienced by the two disks. The phenomenological drag coefficients $\tau_1^{-1}$ and $\tau_2^{-1}$ depend on the sizes of the disks. The internal tension forces $\vec F_a^{(2)\to(1)}$ and $\vec F_a^{(1)\to(2)}$ are Lagrange multipliers that preserve the inextensibility constraint $| \vec R_a^{(1)} - \vec R_a^{(2)} | = \ell$.

The timescales $\tau_1$ and $\tau_2$ control the relaxation of the particle speed towards $\hat{\vec p}_a$. As we did it in Appendix~A, we assume that they are much faster than the evolution of the force field $\vec h$. In this overdamped limit, the dynamics of the dumbbell becomes:
\begin{align}
	&\partial_t \vec R_a^{(1)} = \hat{\vec p}_a + \tau_1 \vec h(\vec R_a^{(1)} - \vec r_b,\hat{\vec p}_b) + \tau_1 \vec F_a^{(2)\to(1)} ,\\
	&\partial_t \vec R_a^{(2)} = \hat{\vec p}_a  + \tau_2 \vec h(\vec R_a^{(2)} - \vec r_b,\hat{\vec p}_b) + \tau_2 \vec F_a^{(1)\to(2)} .
\end{align}
In addition, the inextensibility condition imposes $( \partial_t \vec R_a^{(1)} - \partial_t \vec R_a^{(2)} ) \cdot \hat{\vec p}_a = 0$. Introducing the center of drag of the dumbbell, $\vec r_a = (\tau_2 \vec R_a^{(1)} + \tau_1 \vec R_a^{(2)})/(\tau_1+\tau_2)$, and  assuming that the variations of the force field $\vec h(\vec r-\vec r_b,\hat{\vec p}_b)$ occur on large length scales compared to the particle size, the equations of motion are:
\begin{align}
\label{dumbbell1}	&\partial_t \vec r_a = \hat{\vec p}_a + \bar \tau \left[ 2\vec h(\vec r_a - \vec r_b,\hat{\vec p}_b) + \vec F_a^{(1)\to(2)} + \vec F_a^{(2)\to(1)}  \right] ,\\
\label{dumbbell2}	&\partial_t \hat{\vec p}_a = ({\mathbb I} - \hat{\vec p}_a \hat{\vec p}_a) \cdot \left[ \alpha \vec h(\vec r_a - \vec r_b,\hat{\vec p}_b) + \beta (\hat{\vec p}_a \cdot \nabla) \vec h \right],
\end{align}
where $\bar \tau = \tau_1 \tau_2/(\tau_1 + \tau_2)$, $\alpha = (\tau_1 - \tau_2)/\ell$ and $\beta = - (\tau_1^2 + \tau_2^2)/(\tau_1 + \tau_2)$.
Eq.~\eqref{dumbbell1} describes the translational motion of the particle. Note that the advection terms are  are subdominant in the small-$\bar \tau$ limit (as we explained it in Appendix~A.
The orientational dynamics of the dumbbell due to interactions with particle $b$ is described by Eq.~\eqref{dumbbell2}. This equation is identical to Eq.~\eqref{EM2} with the effective force
\begin{equation}
	\vec F_{a \to b} = \alpha \vec h(\vec r_a - \vec r_b,\hat{\vec p}_b) + \beta (\hat{\vec p}_a \cdot \nabla) \vec h + \mathcal O(\ell^2 \nabla^2) .
\end{equation}

\vfill


\end{document}